\pdfoutput=1 
\documentclass{JINST}

\usepackage{acronym}
\acrodef{DAC}{Digital to Analogue Converter}
\acrodef{DAQ}{Data Acquisition}
\acrodef{DCS}{Detector Control System}
\acrodef{DUT}{Device Under Test} 
\acrodef{HSIO}{High Speed I/O}
\acrodef{RCE}{Reconfigurable Cluster Element}
\acrodef{ROI}{Region Of Interest}
\acrodef{SOC}{System-On-Chip}
\acrodef{TDAQ}{Trigger and Data Acquisition}
\acrodef{ToT}{Time over Threshold}

\usepackage[euler]{textgreek}

\usepackage{subfigure}

\newcommand{\um}{\textrm{{\textmu}m}}
\newcommand{\uRad}{\textrm{{\textmu}m}}

\title{The FE-I4 Telescope for particle tracking in testbeam experiments}

\author{
  \mbox{M. Benoit$^a$},
  \mbox{J. Bilbao De Mendizabal$^a$},
  \mbox{F.A. Di Bello$^a$},
  \mbox{D. Ferrere$^a$},
  \mbox{T. Golling$^a$},
  \mbox{S. Gonzalez-Sevilla$^a$},
  \mbox{G. Iacobucci$^a$},
  \mbox{M. Kocian$^c$},
  \mbox{D. Muenstermann$^{a,1}$},
  \mbox{B. Ristic$^{a,b}$\thanks{Corresponding author.}~},
  \mbox{A. Sciuccati$^{a,2}$}\\
  \llap{$^a$}DPNC, University of Geneva, 
    24 quai Ernest Ansermet, 1211 Gen\`eve 4, Switzerland,\\
  \llap{$^b$}European Organization for Nuclear Research (CERN), 
    385 Route de Meyrin, 1217 Meyrin, Switzerland\\
  \llap{$^c$}SLAC National Accelerator Laboratory,
     Menlo Park, CA 94025, USA\\
  E-mail: \email{bristic@cern.ch}
}
\footnotetext[1]{Now at Physics Department, 
  Lancaster University, Lancaster, UK, LA1 4YB.}
\footnotetext[2]{Now at European Organization for
  Nuclear Research (CERN), 385 Route de Meyrin, 
  1217 Meyrin, Switzerland}

\abstract{
A testbeam telescope, based on ATLAS IBL silicon pixel modules, has been built.
It comprises six planes of planar silicon sensors with $250 \times 50\,${\textmu m$^2$} pitch, read out by ATLAS FE-I4 chips.
In the CERN SPS H8 beamline ($180\,$GeV \textpi$^+$) a resolution of better than $8 \times 12$\,{\textmu}m$^2$ at the position of the device under test was achieved.
The telescope reached a trigger rate of $6\,$kHz with two measured devices.
It is mainly designed for studies using FE-I4 based prototypes, but has also been successfully run with independent DAQ systems.
Specialised trigger schemes ensure data synchronisation between these external devices and the telescope.
A region-of-interest trigger can be formed by setting masks on the first and the last pixel sensor planes. 
The setup infrastructure provides centrally controlled and monitored high and low voltage power supplies, silicon oil cooling, temperature and humidity sensors and movable stages.
}


\begin{document}

  \section{Introduction}\label{sec:intro}
      Testbeam experiments using charged particle tracking by beam telescopes have proven to be a useful tool for detector characterisation.
Over the last decades increasingly powerful beam telescopes in terms of resolution and data rate have been constructed.
The availability of the FE-I4~\cite{fe-i4} readout ASIC, that has been designed and employed in the ATLAS IBL\cite{ibl}, has allowed the construction of a telescope to test small prototypes with trigger rates of the order of several kHz for devices that cover surfaces as small as a few mm$^2$, with a pointing resolution at the \ac{DUT} of typically $10\,$\um.

The six IBL telescope modules, combined with centrally managed services, comprised of power supplies, cooling by means of a silicon oil chiller and DUT positioning using XY-stages, provide a fully integrated system for silicon sensor characterisation in high energy beams  like at the CERN SPS H8 ($180\,\textrm{GeV}\,\pi^+$).

This paper presents a description of the hardware and software of the setup, as well as of the results in terms of resolution at the \ac{DUT} position and data acquisition rate capability.

\newpage

  \section{Telescope description}\label{sec:desc}
      \begin{figure}[ht]
  \centering
  \includegraphics[width=.95\textwidth]{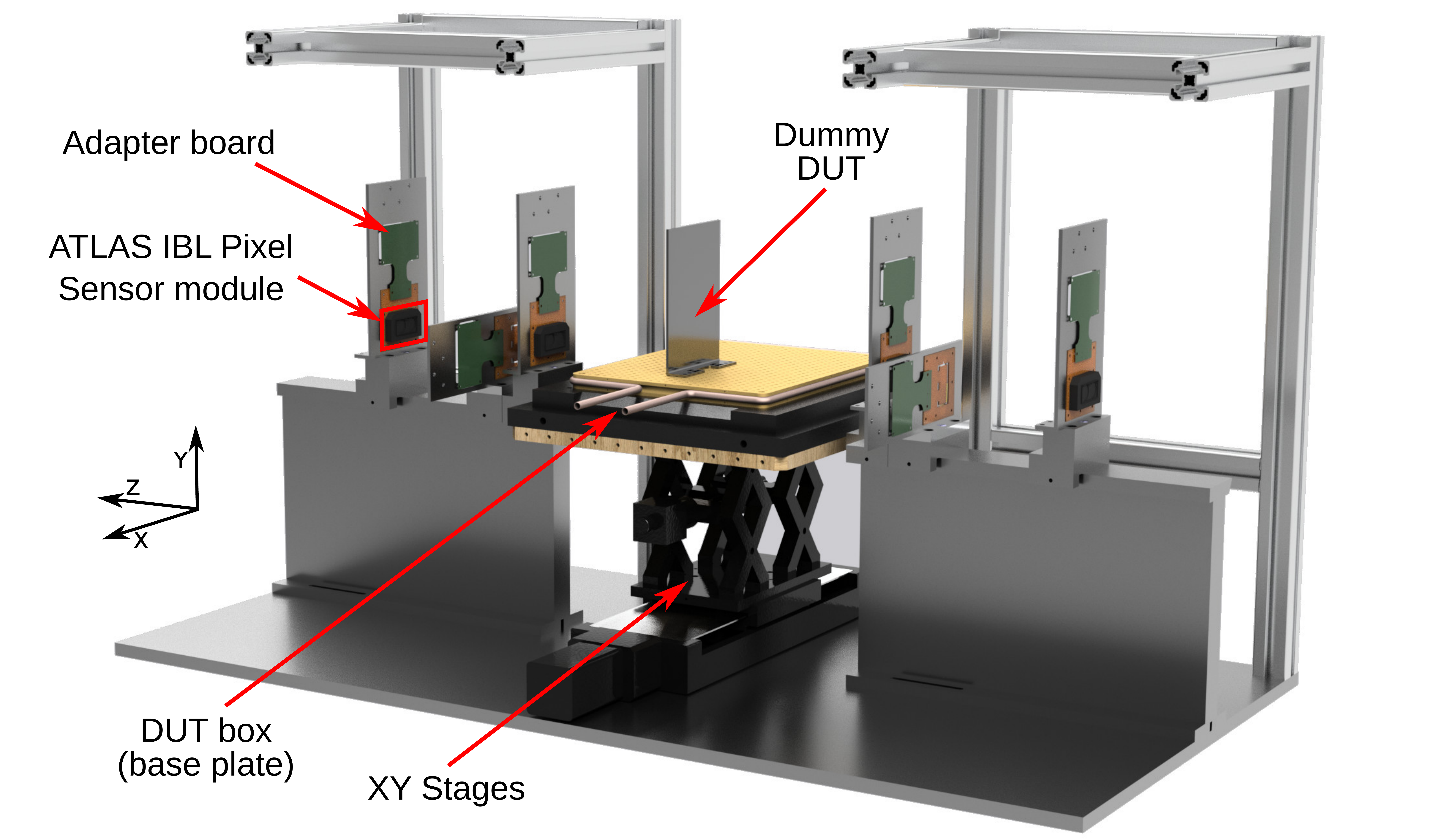}
  \caption{\label{fig:tel3d}Computer generated image of the telescope setup, showing the two telescope arms holding the six FE-I4 pixel telescope planes, the XY stages with the base plate of the DUT box and a dummy DUT.}
\end{figure} 
The FE-I4 telescope, shown in Figure \ref{fig:tel3d}, consists of two arms, each hosting three detector planes equipped with a $200\,\um$ thick n$^+$-in-n planar pixel silicon sensor~\cite{ibl} with a pixel pitch of $50\times250\,\um^2$.
As the sensor pixels are not square shaped, the middle plane of each arm is rotated by $90^\circ$ around the beam axis to ensure a comparable spatial resolution in the X and Y directions.

Each sensor is read out by two FE-I4 chips, of which only one is active for data taking.
The FE-I4, built in a $130\,$nm CMOS process, hosts a pixel matrix of 336 rows by 80 columns forming an active area of $20.2 \times 16.8\,\textrm{mm}^2$.
Each pixel is DC coupled to a two stage amplifier and shaper which is followed by a discriminator.
Thresholds can be set globally with local corrections applied per pixel by \acp{DAC}.
The deposited charge is stored as a 4-bit \ac{ToT} signal clocked at $40\,$MHz, the nominal bunch crossing frequency of the LHC.
Detected hits and 8-bit timestamps are stored in memory cells shared by four pixels, that are read out following a trigger signal.
Each memory cell can store up to 5 events, which are retained for an externally defined latency interval and then erased by further hit information.
This design allows for a hit rate of up to $400\,$MHz/cm$^2$.

The assemblies of FE-I4 chips and sensors are mounted on aluminium frames that also provide passive cooling for the detectors.

The electrical connection to each assembly is made via a flexible PCB that is glued to the backside of the sensor.
It hosts the necessary wire bond pads and noise filtering circuitry, and connects the assembly to an adapter board mounted on top of the aluminium frame.

\vfill

\subsection{DAQ system and software}
\label{sec:daq}
Three components form the \ac{DAQ} system; the \ac{HSIO} board; the \acp{RCE}~\cite{rce}; and a dedicated PC running the ATLAS \ac{TDAQ} framework.

The \ac{HSIO} is a custom-made general-purpose readout board based on a Xilinx Virtex-4 FPGA.
It hosts a large number of I/O channels and receives (sends) data from (to) the higher level processing \acp{RCE}.
It implements the command and data protocols of the FE-I4, relays commands from the \ac{RCE} and buffers data from the front-end.
It also generates the clock, that is provided globally to all detectors.
The \ac{HSIO} issues triggers to the telescope and DUT by evaluating multiple configurable inputs.
In the case of the telescope, two planes can be configured to output a logical OR of all or a subset of pixel discriminators, that is then directly fed into the FPGA.
This permits to define a region of interest trigger, which is especially useful for devices smaller than the telescope's acceptance.
A trigger veto can also be supplied externally for running independent \ac{DAQ} systems in parallel with the telescope DAQ (see section \ref{sec:dut}).
The data are sent via optical fibers to as many as four \ac{RCE} computing units with a rate of $3.125\,$Gbps per fiber.

\ac{RCE} is a generic computational unit based on a \ac{SOC} which can handle up to 24 lanes of high-speed serial I/O at lane speeds up to more than $40\,$Gb/s.
The core is a $350\,$MHz PowerPC processor running an RTEMS Real-Time kernel.
The \ac{RCE} configures the \ac{HSIO} and sends commands to the interfaced front-ends.
Raw front-end data coming from the \ac{HSIO} is histogrammed and transferred via Ethernet to the \ac{DAQ} PC.

The DAQ PC runs a Linux OS and the ATLAS \ac{TDAQ} system for communication and data handling.
Custom-made software allows for setting per-plane configurations as well as global parameters such as the trigger mode and delays.
The trigger can be set to evaluate a combination of digital and discriminated analogue signals.
A configurable cyclic trigger is available for debugging purposes.
Trigger delays, front-end latency values and readout window lengths can be set separately for the telescope planes and DUT to accommodate for possibly different response times of the controlled devices.
The settings are accessible via a simple GUI and stored in text configuration files.

During data taking an online monitor provides real-time hit maps, correlation plots between the hit positions of neighbouring planes, and timing and charge information.

  \subsection{Mechanics and services}\label{sec:mec}
      \begin{figure}
  \centering
  \includegraphics[width=.7\textwidth]{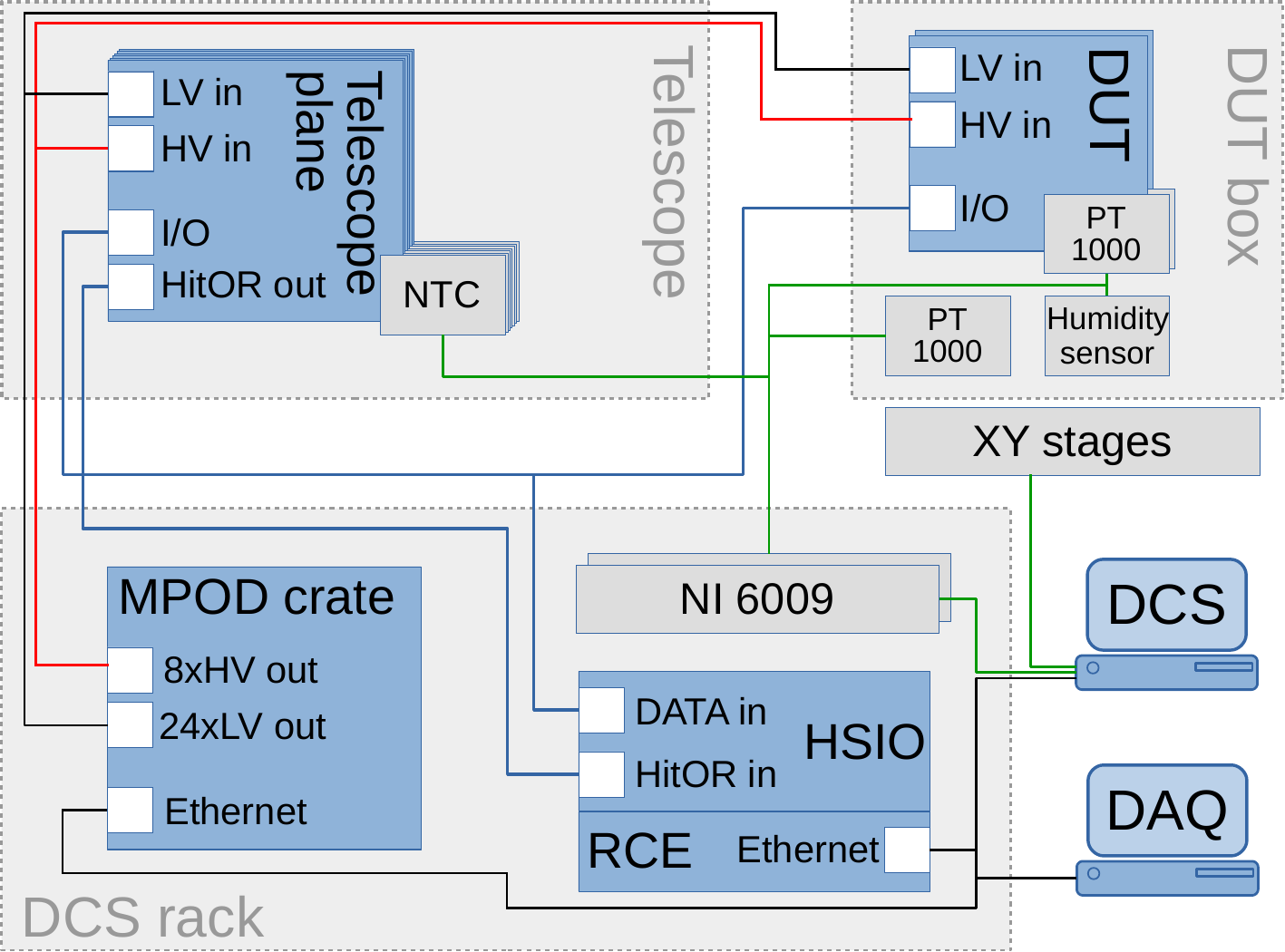}
  \caption{\label{fig:tel_scheme}Schematic view of the telescope setup including the DUT and the services.}
\end{figure} 
The testbeam setup as sketched in Figures \ref{fig:tel3d} and \ref{fig:tel_scheme} consists of two mechanical structures, the telescope itself and a movable rack containing the \ac{DAQ} and \ac{DCS} components.
The telescope is composed of a baseplate holding the two telescope arms, the linear stages and the \ac{DUT} box.
Each of the three telescope planes in each arm can be positioned freely along the $40\,$cm length telescope arm and tilted around the short pixel direction in order to adapt to different requirements given by the beam and telescope user.
Both telescope arms can be moved separately to accommodate DUT setups with a width between $20$ and $40\,$cm, thus ensuring a positioning of the innermost planes as close as possible to the DUT.

The test-box containing the DUT is accurately positioned in the XY-plane (perpendicular to the direction of the beam) by means of two high-precision linear stages equipped with stepper motors ($1.8^\circ$ rotor tooth-pitch).
The stages have a travel range of 50 mm and 110 mm respectively along the horizontal (X) and vertical (Y) directions.
An OES ALLEGRA-2-10-02~\cite{oes} two-axis programmable control system driver is used to control the stages via a RS-232 serial port.
The driver provides up to 256 micro-steps per step to increase the motor resolution.
The absolute positional accuracy is $\sim 2\,\um$ (for the typical case of 10 micro-steps per step).
Single-pole single-throw switches, located at both ends of each axis, and connected to the motor driver, are used as an interlock safety system, for automatically stopped motion when reaching any axis limit.

\begin{figure}[tbp]
  \centering 
  \includegraphics[width=.9\textwidth]{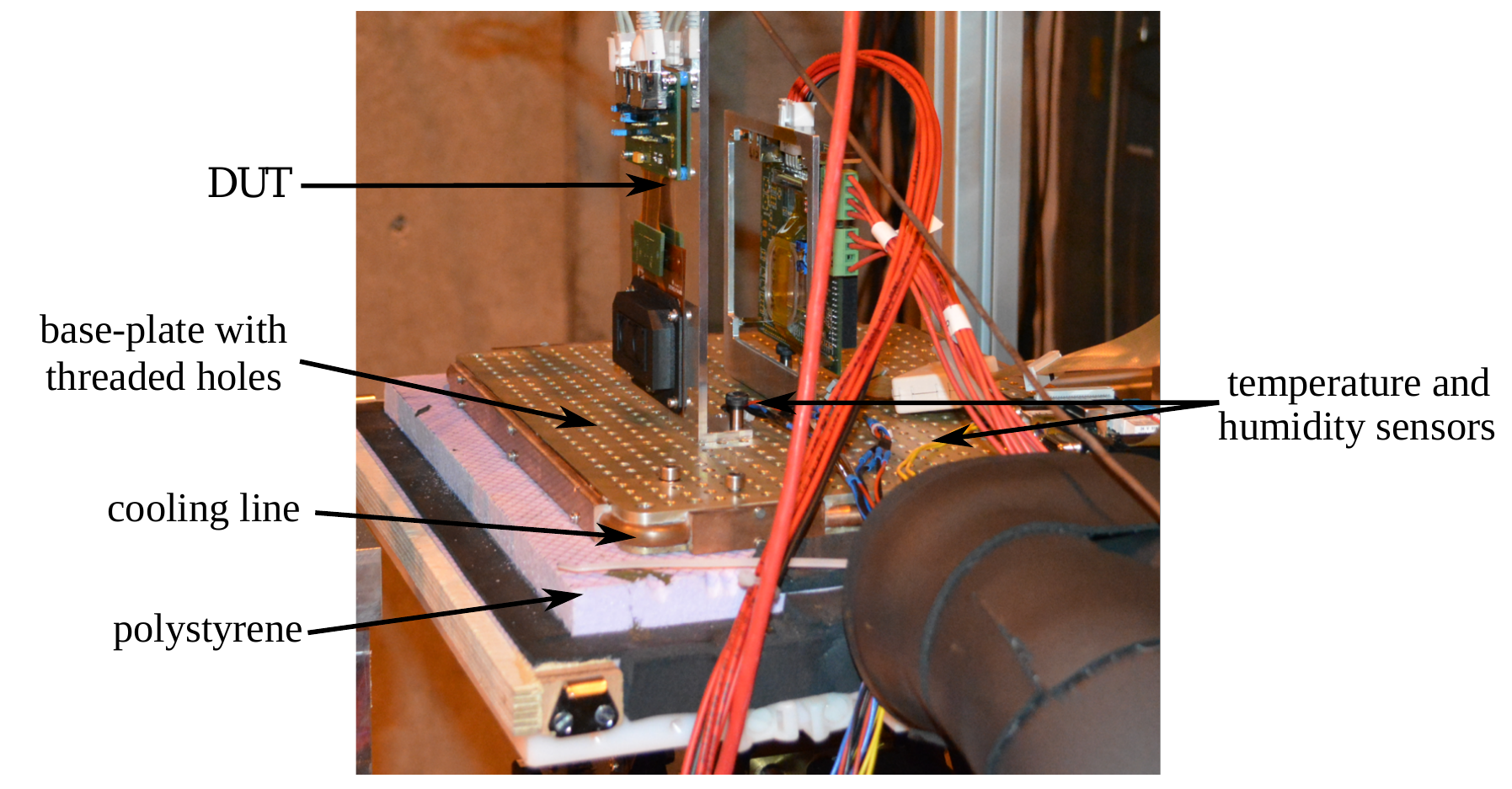}
  \caption{\label{fig:tel_cooling}Photograph of the test-box with two DUTs installed.
Only the bottom-side of the box is shown.
The inlets for the nitrogen flux located in the top-side of the box-lid are not visible.}
\end{figure}

\paragraph{Power supply}
\label{sec:ps}
The different voltages needed to operate the telescope modules and the DUTs are provided by a set of low-voltage (LV) and high-voltage (HV) power supply (PS) modules installed in an MPOD mini-crate manufactured by WIENER~\cite{wiener}.
The MPOD mini-crate is a compact 19'' rack-mountable chassis that has been equipped with four eight-channel PS cards (6U Eurocard format) in a mixed configuration: one ISEG EHS-8210n-F~\cite{iseg} HV module for the bias of the telescope planar sensors, two WIENER MPV 8008D LV modules and a WIENER MPV 8210D LV module for the voltage supply of the readout ASICs and any other additional voltages.
Each HV channel has a floating ground and can provide up to $-1000\,$V and maximum $8\,$mA current through isolated built-in SHV connectors.
The two different types of LV modules conveniently cover the range of low (up to $8\,$V, maximum current of $10\,$A) and medium (up to $120\,$V, maximum current of $100\,$mA) voltages typically needed, for example to bias the DUT sensor.
The signals of four LV channels are driven through a 37-pin sub-D connector.
For each channel, both DC and sense measurements are provided.
Finally, an integrated MPOD crate-controller manages the primary power supplies and provides 10/100 Ethernet, CAN bus and USB 2 interfaces for remote monitoring and control.
Local control is performed via an LCD display and two rotary controls.
\newpage
\paragraph{Cooling}
\label{sec:cooling}
The six telescope planes are operated at room temperature without need of active cooling.
The DUTs are mounted in a light-tight thermally insulated test-box.
A dry atmosphere is ensured by flushing cold nitrogen gas inside the box.
A $30\times25\,\textrm{cm}^2$ aluminum base-plate with a grid of M5 holes with $1\,$cm pitch (see Figure~\ref{fig:tel_cooling})
is used to position the DUT holding-frames and to ensure a good cooling contact surface.
A copper cooling pipe running laterally along the sides of the base plate is connected to a silicon-oil HUBER UNISTAT 705~\cite{huber} chiller that can reach a minimum temperature of $-75\,^\circ$C.
The chiller can be controlled remotely via its RS-232 / RS-485 serial socket.

\paragraph{Temperature and humidity monitoring}
\label{sec:tempmonitoring}
The temperature of each telescope module, measured through a NTC attached to the corresponding flex-hybrid, is continously recorded by a National Instruments (NI) USB-6009 OEM~\cite{ni} multi-channel (14-bits resolution) USB DAQ device.
A second NI DAQ device is used to monitor the environmental conditions by reading additional PT1000 temperature and humidity sensors located at several locations inside the test-box.
A few readout channels are reserved for reading the temperature sensors located close to the DUT.

\begin{figure}[tbp]
  \centering 
  \includegraphics[width=.8\textwidth]{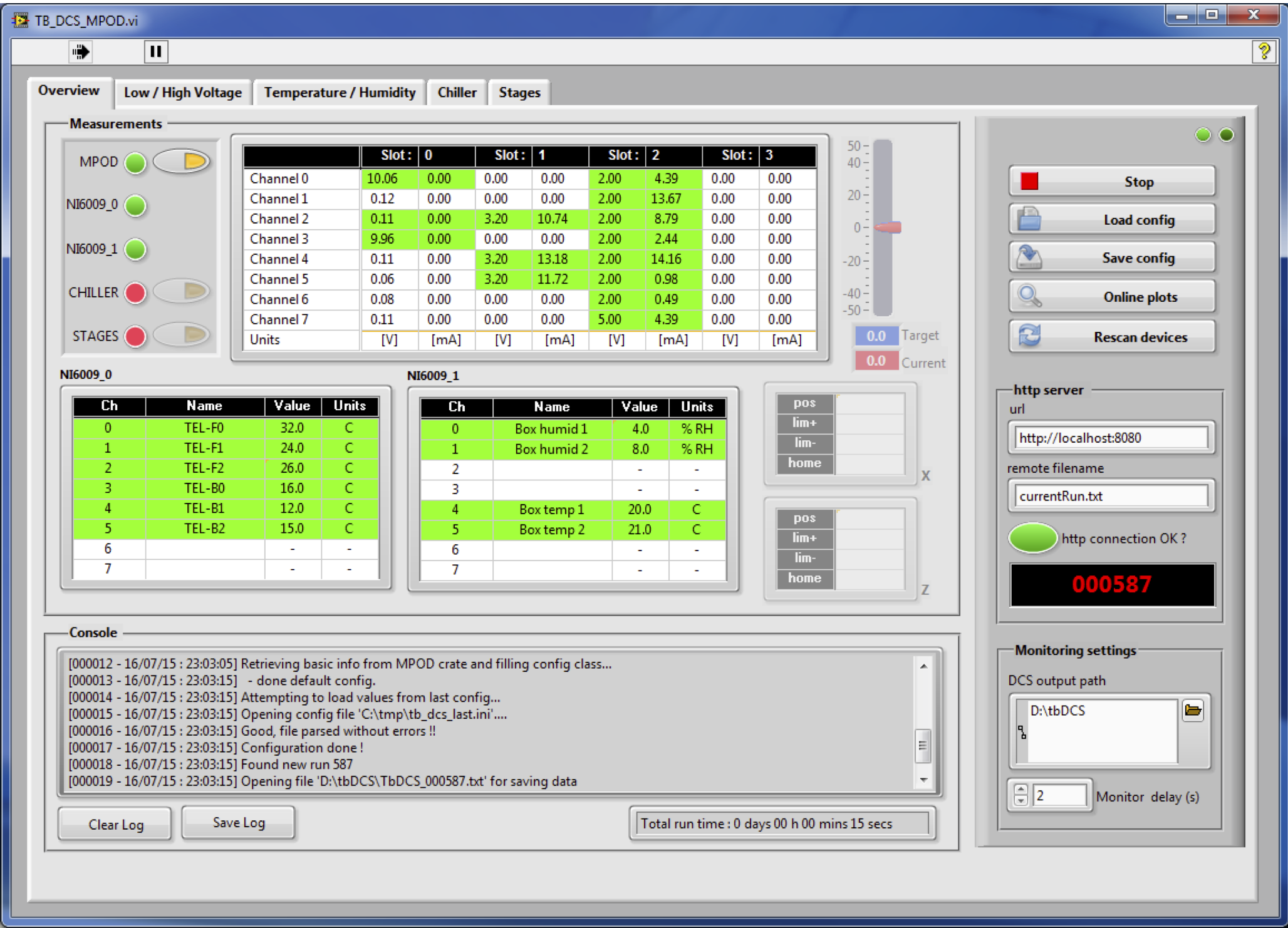}
  \caption{\label{fig:lv_dcs}Graphical User Interface of the NI LabVIEW slow control software.
The {\em Overview} tab shown provides a summary of all the different data measurements and the device status}
\end{figure}

\paragraph{Slow control software}
\label{sec:slow_control}
A single host computer is used to control the MPOD mini-crate, the linear stages and the chiller, as well as to record the various temperature and humidity data.
The slow control software has been developed using the NI LabVIEW 2012 Professionnal Development System in an object-oriented programming approach.
The program GUI is shown in Fig.~\ref{fig:lv_dcs}.
Upon startup, the slow control software establishes an HTTP connection to the DAQ PC to access the current run number.
The communication with the MPOD mini-crate is made using the SNMP protocol.
For the PS modules, the user has full-control over the most commonly used functionalities, including per-channel current limits, voltage-ramp speeds, etc.
In case a specific power-up order is needed, single supply channels can be grouped in an arbitrary way in so-called ``channel-sequences'', which includes the possibility of mixing channels from different PS modules.
Upon start of a channel-sequence, the list of channels is automatically run through with a user-defined delay between the voltage of two consecutive channels.
Various program parameters ({\em e.g.} channel names, PS current limits and composition of channel-sequences) can be defined in text configuration files.
An output file is created for each  run.
Data is recorded in a simple ASCII format with a user-defined delay, typically set to a few seconds.

  \subsection{DUT integration}\label{sec:dut}
      The FE-I4 telescope was initially designed to test devices read out by the same FE-I4 chip, as it is widely used by the ATLAS Pixel community for sensor R\&D.
Therefore FE-I4 based DUTs are read out exactly like telescope planes.
To cope with different characteristics of the DUTs, the readout window and delay settings can be set separately for the telescope and DUT planes.

\begin{figure}[ht]
  \centering
  \includegraphics[width=.8\textwidth]{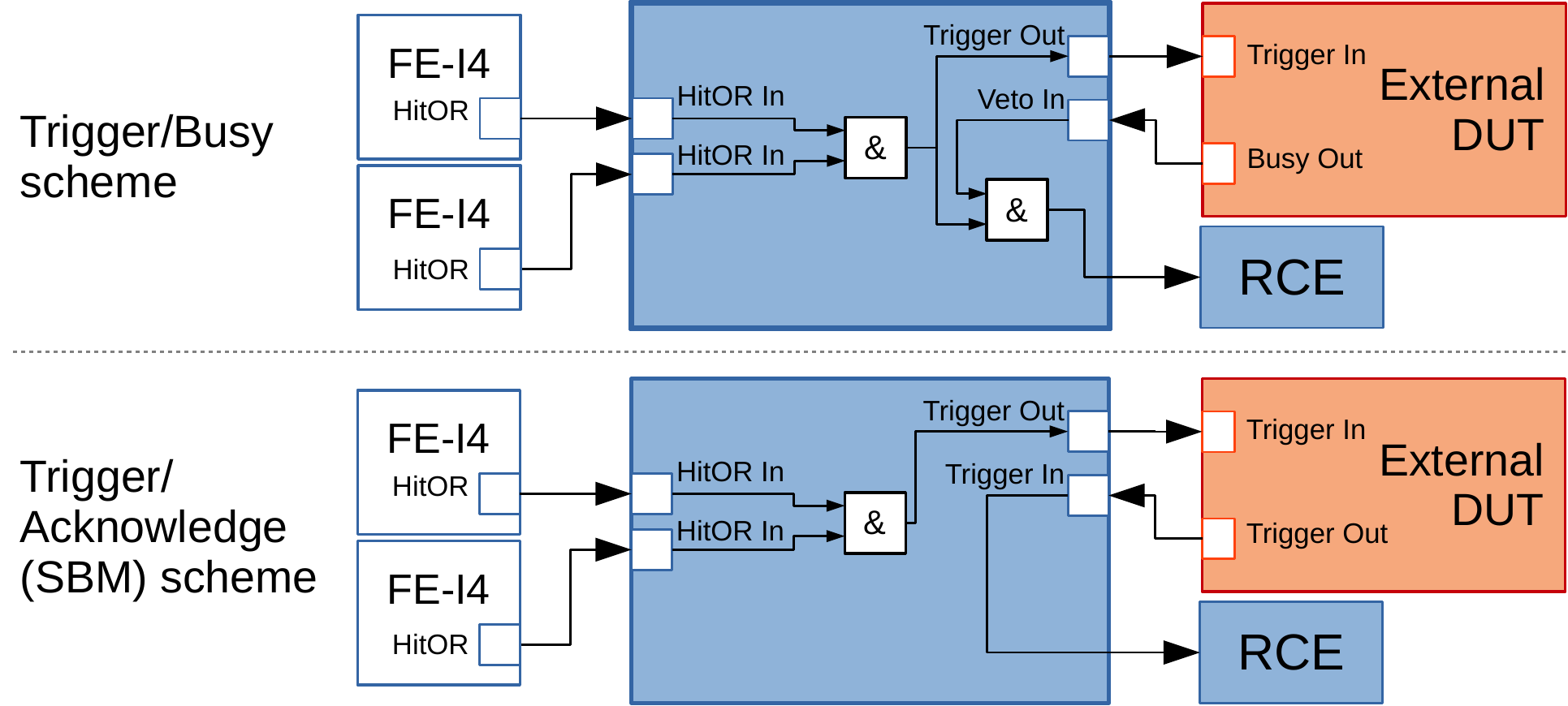}
  \caption{\label{fig:trig_scheme}Trigger schemes for non-FE-I4 (external) DUTs ensuring synchronised data streams which can be merged off-line for analysis.}
\end{figure}

Non-FEI4 types of DUTs (``external'' DUTs) are not implemented in the RCE readout and have to rely on their own readout system.
There are two schemes for synchronising the telescope and the external DUTs as seen in Figure \ref{fig:trig_scheme}.

In the first scheme (\emph{Trigger / Busy scheme}) the telescope triggers itself and sends a TTL pulse to the external DAQ system that then issues a busy signal to the telescope.
During the busy signal further telescope triggers are ignored.

In the second scheme (\emph{Trigger / Acknowledge scheme}), the trigger signal generated by the telescope is delivered directly to the external DAQ system.
If this system is ready for data taking and accepts the trigger it sends a signal to the RCE to read out the telescope.
Both schemes ensure synchronised data streams that have to be merged off-line for analysis.
\vfill

  \section{Reconstruction}\label{sec:rec}
\begin{figure}[ht]
  \centering
    \includegraphics[width=.53\textwidth]{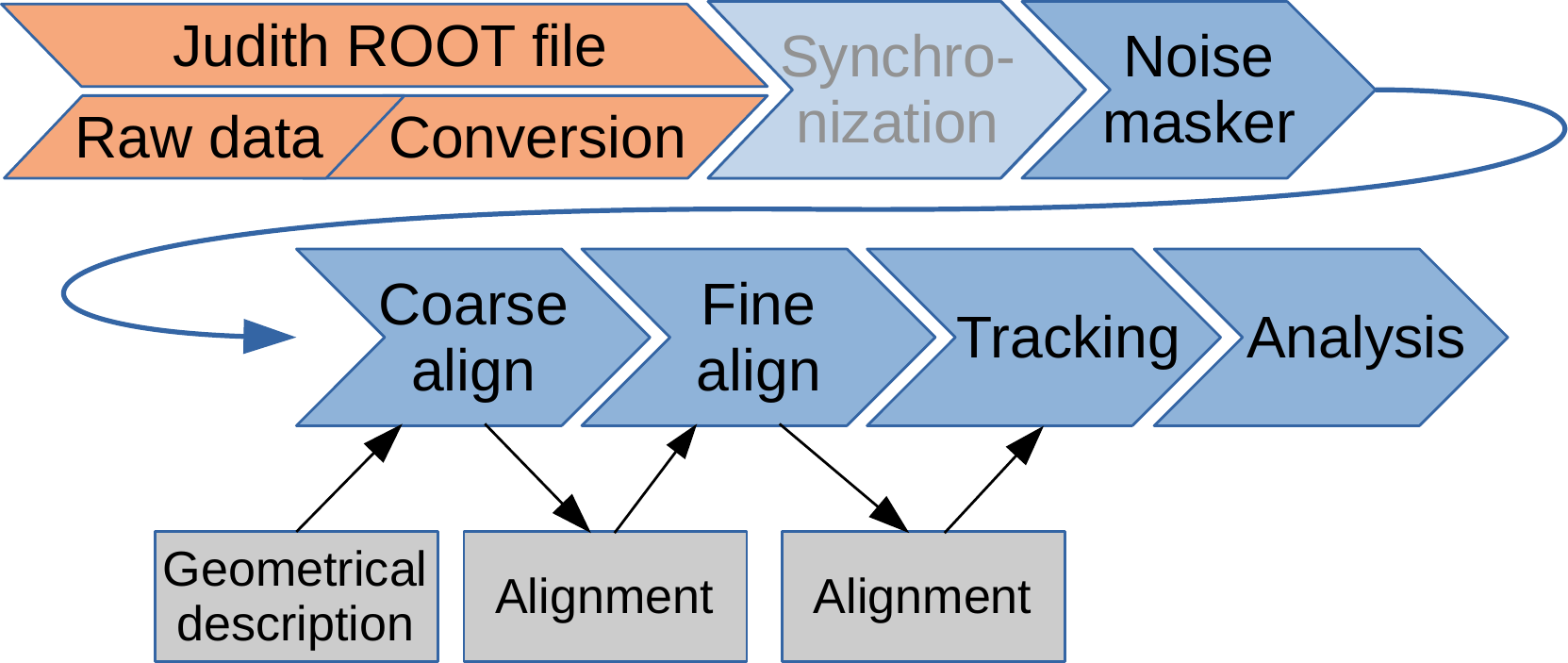}
  \caption{\label{fig:judith_flow}Data flow in the Judith reconstruction framework. The synchronisation step is only needed when external DUTs miss triggers, thus desynchronize, which has not been observed with this telescope so far.}
\end{figure}

      The raw telescope data is handled by the Judith~\cite{judith} reconstruction software.
Judith is a fully object oriented framework written in C++, aiming for speed and simplicity.
It is therefore composed as a collection of specialised algorithms for data handling and synchronisation, setup description, reconstruction and analysis.
The raw data has to undergo several processing steps in order to be available for analysis, as depicted in Figure \ref{fig:judith_flow}.
All data are stored in ROOT~\cite{ROOT} file format, with the geometry description and configuration of the framework given in text configuration files.

\subsection{Preprocessing and clustering}
The first step is a noise masking algorithm that calculates the per pixel occupancies and excludes from further analysis those pixels which exceed a certain counting threshold.
Judith also implements a sophisticated synchronisation algorithm, which so far was not necessary for the data collected with the telescope described in this paper.

Particles traversing a detector plane can trigger more than one pixel, leading to clusters of pixels.
Judith therefore translates hits into clusters before further processing.
Beginning with random seed hits, a recursive algorithm groups all neighbouring hits.
Overlapping clusters cannot be detected and are considered as one.
The actual position of the hit is then calculated as the geometric mean of the pixel positions in the cluster.

\subsection{Alignment}
In order to reconstruct particle trajectories with high precision, the relative position between detector planes of the testbeam setup has to be accurately known.
Starting with the positions and rotations of the planes measured by the telescope survey, a two step alignment procedure is applied to correct for lateral misplacement and rotations around the Z-axis.

A first coarse alignment calculates 2D inter-plane hit correlation distributions, which in case of a parallel beam should yield a straight line.
Given a perfect alignment the line would cross the origin with a slope equal to the ratio of the pixel pitches.
Shifts between the planes are therefore deduced from the offset of the line from the origin and rotations from the deviation of the measured to the calculated slope.
The first plane is fixed as point of reference and planes are only aligned to their direct neighbours to minimise the influence of scattering.

In a second step, a fine alignment calculates track residuals for precise positioning of the planes, as the distance between the projected track position and the position of the associated cluster on a plane.
The correction to the position of the plane is then calculated from the residual and the hit position on that plane.
As the track position is directly influenced by the current alignment of the telescope the plane under investigation is always excluded from track fitting, forming unbiased tracks.

\subsection{Track reconstruction}
Starting from a seed cluster, the track reconstruction algorithm searches for clusters on consecutive planes within an user-defined cone angle.
The angle is chosen according to the scattering angle on the planes, so that scattered clusters are included.
The algorithm initially assumes that the track is parallel to the telescope's longitudinal direction, but this is modified as clusters are added to the track.
If multiple candidate clusters are found on a plane, the track bifurcates and further searches continue for each cluster.
Candidate tracks with the largest number of clusters are kept, straight-line fits are performed and the one with the smallest $\chi^2$ is selected.
Clusters that have been assigned to the selected track are excluded from further searches.
The algorithm copes well with large scattering angles given a low track density, thus a large enough separation of tracks, so clusters are correctly assigned to their corresponding track.
The fit $\chi^2$ is then used to filter those tracks which are disturbed largely by scattering or nuclear interaction with the telescope planes.

  \section{Performance}\label{sec:per}
      The FE-I4 telescope was characterised at the CERN SPS H8 beamline ($180\,$GeV \textpi$^+$) in 2014 and 2015.
During multiple testbeam periods, full size FE-I4 as well as small-area ($3\,\textrm{mm}^2$) sensors were tested.

Sample data sets of several million triggers were used for analysis.
For $98.7\%$ of events, tracks were reconstructed with $96.1\%$ being single track and $3.7\%$ double track events.
Up to eight tracks were built per event.
Although the SPS aims for providing a mostly constant beam intensity of typically some $100\,$kHz during a spill, high intensity bursts can occur.
The bursts and particles from nuclear interaction on the instrumentation in the beam line and the telescope are likely to account for the high track number events.
Due to the time resolution of the FE-I4 chips of $25\,$ns, particles in such events cannot be differentiated by time of arrival.
Tracks which pass close to each other therefore cannot be reconstructed reliably and are discarded.

A typical requirement on the tracks for analysis is the detection of the traversing particle by all six telescope planes, which is the case for $96\%$ ($3.2\%$ formed by five clusters) of the reconstructed tracks.
This leads to an average detection efficiency per plane of $99.4\%$.

\subsection{Trigger rate}
The trigger performance of the telescope is summarised in Table \ref{tab:perf_trig}.
In standalone mode, the telescope records over 18k triggers per second, which defines the maximum data taking rate for external devices.
Higher trigger rates are in principle possible but at the expense of excluding telescope planes, thus a reduced spatial resolution.
For one or two FE-I4 based DUTs the trigger rate was found to remain stable at \textasciitilde$6\,$kHz.
Small devices that geometrically cover only a fraction of the telescope's acceptance profit greatly from the \ac{ROI} trigger.
In case of a $3\,\textrm{mm}^2$ device, the fraction of recorded tracks passing the DUT is increased from 2\% to about 70\%.
Currently the data rate is limited by the DAQ system which will be improved in future developments.
\begin{table}[hptb]
  \caption{\label{tab:perf_trig}Trigger performance of the telescope setup for different configurations: standalone, with FE-I4 and small sized DUT. The telescope planes readout window was fixed at $100\,$ns.}
  \begin{center}
    \begin{tabular}{ccccc}
      \hline
      Number of              & DUT readout & ROI      & Trigger rate & Fraction of \\
      DUT                    & window      & trigger  & [kHz]        & in-DUT tracks \\ \hline
      none                   & /           & no       & 18.3         & \\
      1 ($3\,\textrm{mm}^2$) & $400\,$ns   & no       & 6.3          & \textasciitilde2\%\\
      1 ($3\,\textrm{mm}^2$) & $400\,$ns   & yes      &	6.0          & \textasciitilde70\%\\
      1 (FE-I4 size)         & $200\,$ns   & no       & 5.7          & >90\%\\
      2 (FE-I4 size + $3\,\textrm{mm}^2$) & $400\,$ns & no & 6.3     & >90\% / \textasciitilde2\%\\\hline
    \end{tabular}
  \end{center}
\end{table}

\subsection{Spatial resolution}
\label{sec:spat}
The spatial resolution of the telescope is estimated by the error of the track fit parameters projected to the position of the DUT.
Due to the digital hit information available (no charge information is taken into account), the resolution only depends on the positions of the planes and on the telescope pixel granularity, and is calculated to be $\sigma_X=11.7\,\um$ and $\sigma_Y=8.3\,\um$.

The telescope plane material budget comprises of $350\,\um$ of silicon, a $100\,\um$ thick aluminium holding structure at the edges of the detectors and ABS covers to shield the detectors from the environment.
Hence the multiple scattering effect was estimated in tracking simulations for an $180\,$GeV \textpi$^+$ beam.
A scattering angle of a seven plane telescope (the middle one acting as DUT) of $14.76\,\uRad$ ($6.57\,\uRad$ per plane) was calculated, which leads to a deflection of $\sim12\,\um$ over the $80\,$cm long telescope.
In view of this result the calculation was repeated for a four plane setup, with every second plane rotated by $90^\circ$.
The total scattering angle of $14.68\,\uRad$ is comparable to the 6 planes telescope, whereas the resolution at the telescope centre decreases by $1\,\um$ in the X and $3\,\um$ in the Y direction.

\begin{figure}[ht]
  \begin{center}
    \subfigure[]{
      \includegraphics[width=0.475\textwidth]{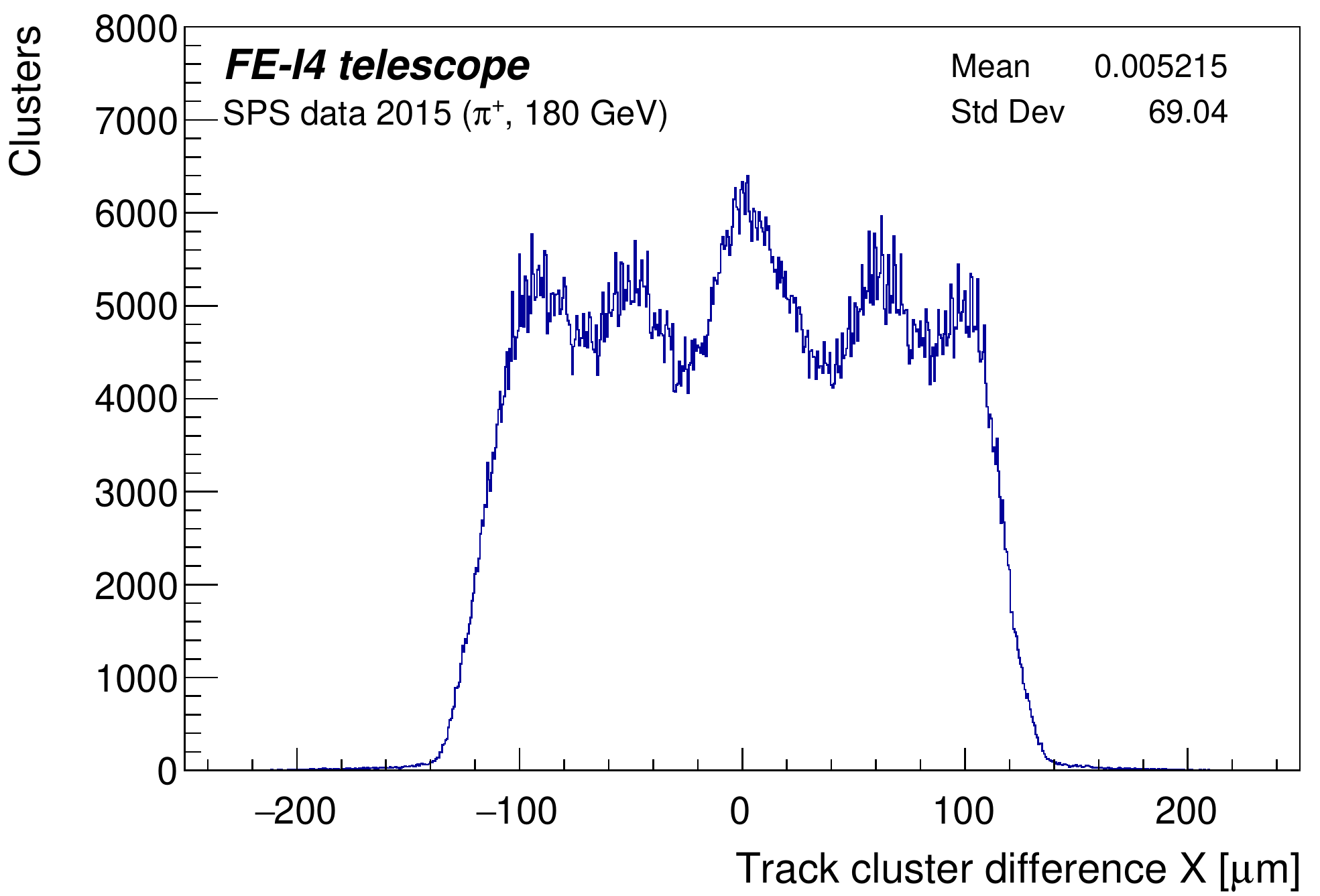}
      \label{fig:residualX}
    }
    \hfill
    \subfigure[]{
      \includegraphics[width=0.475\textwidth]{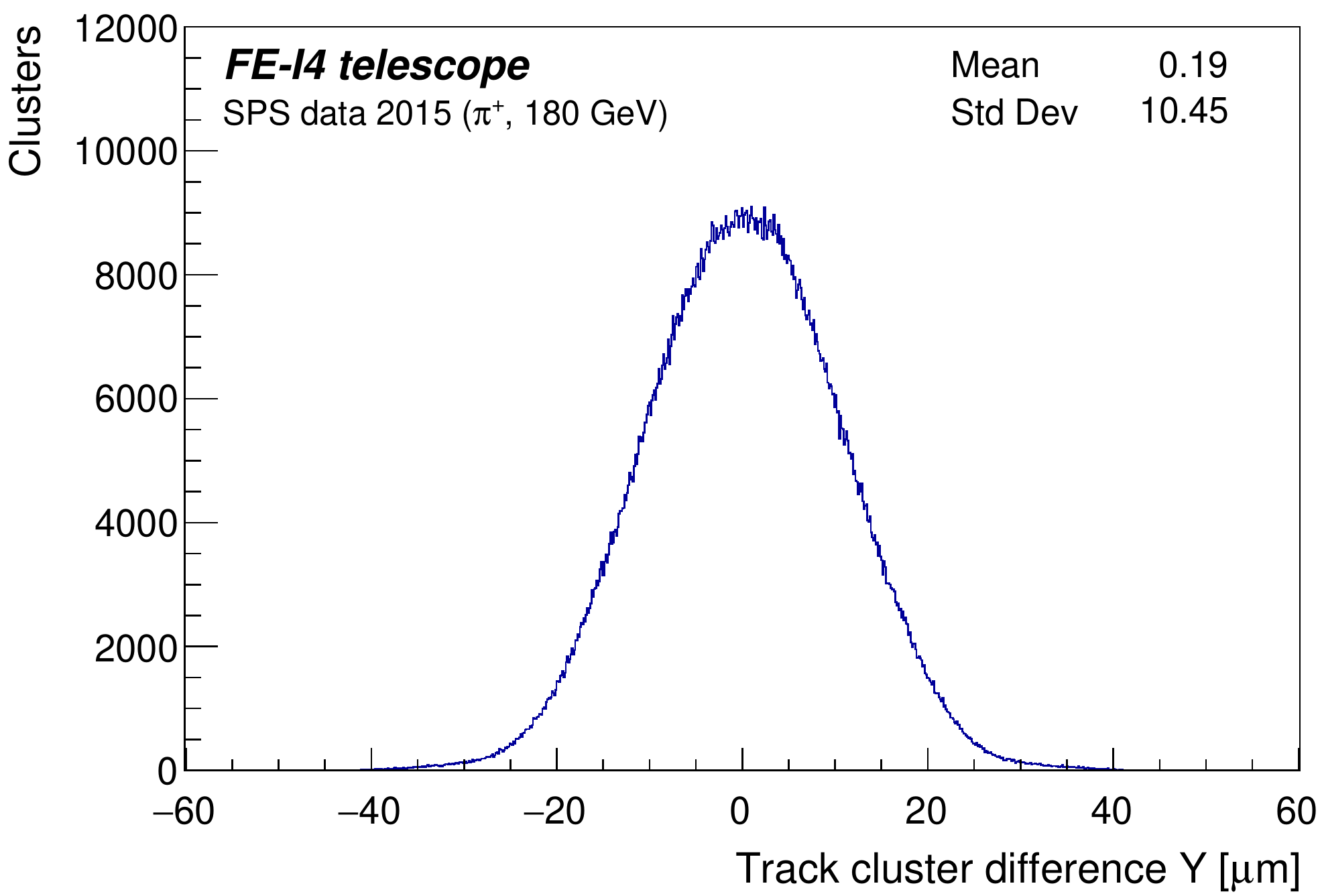}
      \label{fig:residualY}
    }
    \vspace{-6pt}
    \caption[]{\label{fig:residuals} Residuals distribution in \subref{fig:residualX}~X and \subref{fig:residualY}~Y directions for the third telescope plane after alignment.} 
  \end{center}
\end{figure}

Figures \ref{fig:residualX} and \ref{fig:residualY} show the residual distributions for the  telescope plane that is closest to the DUT box (third plane) in the X and Y directions after alignment.
While the residuals in Y resemble the expected shape of a Gaussian distribution (actually, an uniform distribution convoluted with a Gaussian distribution), those in X direction show a five-peak structure.
This is a geometric effect, as it will be shown below, that results from the increased granularity given by the rotated telescope planes.

\begin{figure}[!ht]
  \begin{center}
    \subfigure[]{
      \includegraphics[width=0.475\textwidth]{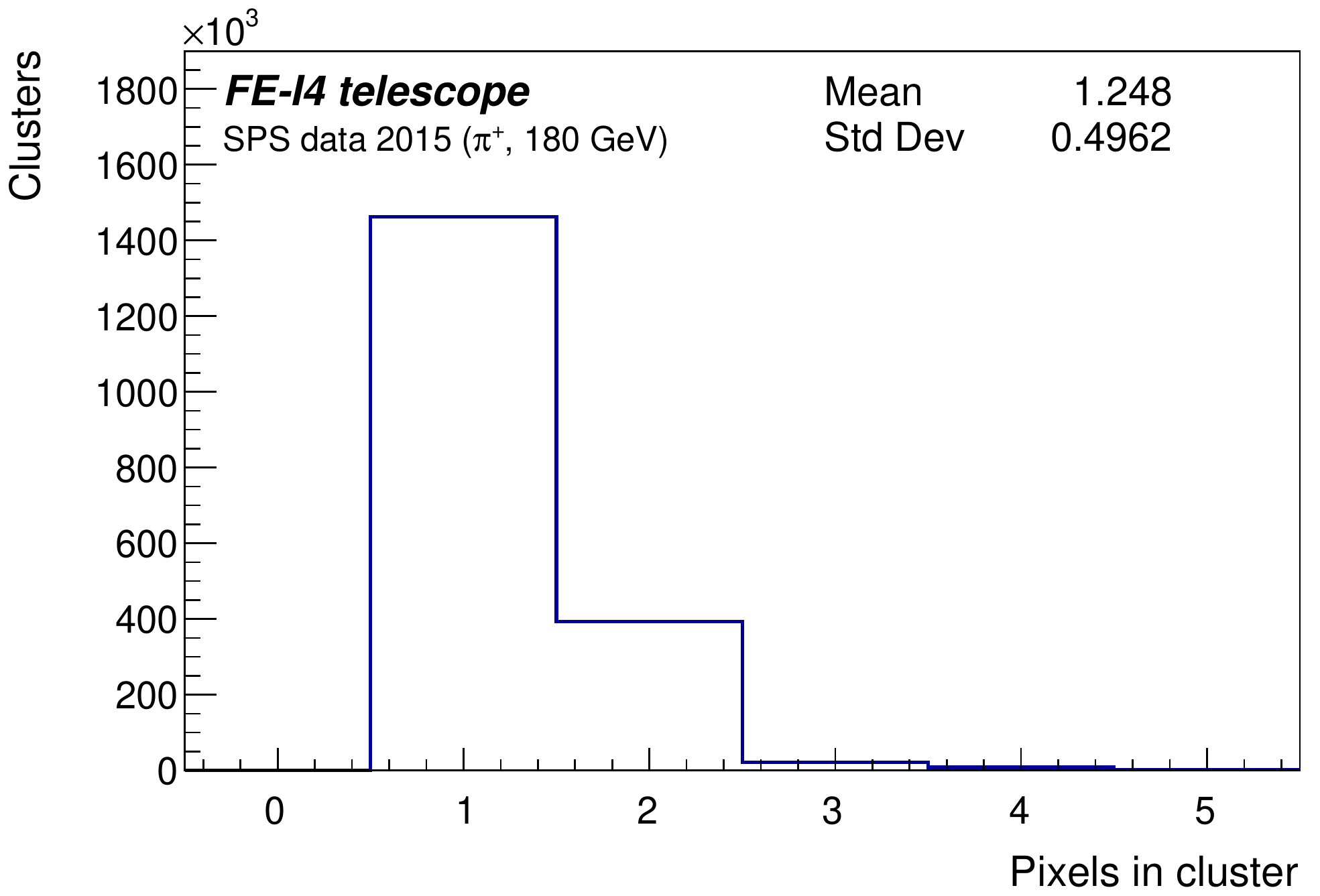}
      \label{fig:clusize}
    }
    \hfill
    \subfigure[]{
      \includegraphics[width=0.475\textwidth]{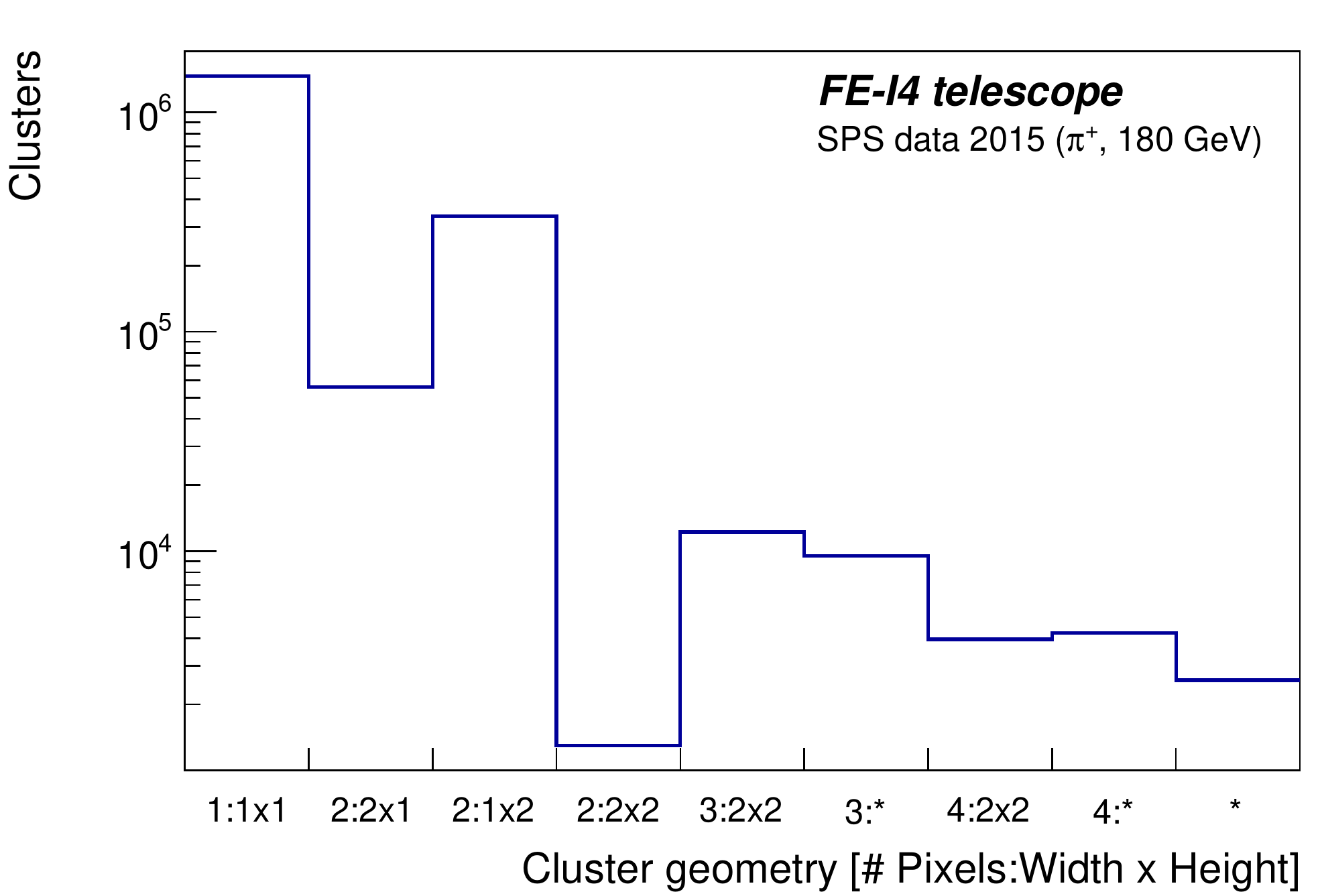}
      \label{fig:clugeom}
    }
    \vspace{-6pt}
    \caption[]{Distribution of \subref{fig:clusize} cluster sizes and \subref{fig:clugeom} shapes of clusters for the third telescope plane. Cluster shapes, that are not explicitly listed for a given number of pixels, are merged in one bin, labelled with an asterisk.} 
  \end{center}
\end{figure}

Both distributions are centred around zero with a standard deviation (SD) of $\sigma_{X,res}=69.0\,\um$ and $\sigma_{Y,res}=10.5\,\um$.
The SD of the residual distribution can be calculated with $s_i=p_i/\sqrt{12}$ being the resolution of the sensor and $p_i$ the pixel pitch in X or Y direction.
The calculated values of $\sigma_{X,res}=73.1\,\um$ and $\sigma_{Y,res}=16.7\,\um$ are larger than those measured.
This is attributed to the charge sharing effect, which for the telescope sensors leads to around 20\% of clusters containing more than one hit as shown in Figure \ref{fig:clusize}.

The most-probable area for a particle to produce a one pixel cluster is smaller than the actual pixel size.
This leads to a reduced effective pixel size, thus an increased resolution for single hit clusters, which is also reflected in the SD of their residual distributions $\sigma_{X,res,SP}=70.2\,\um$ and $\sigma_{Y,res,SP}=10.6\,\um$.
The low threshold and large signal produced by the planar sensors makes it improbable for a particle to fire less than two pixels when hitting within a zone of around $7\,\um$ in X and $14\,\um$ in Y around the perimeter of the pixels.
This asymmetry is also supported by the distribution of cluster geometries as shown in Figure \ref{fig:clugeom}.
Therefore the resolution of single sensors is overestimated and by this the given telescope resolution is merely an upper limit.

\paragraph{The five peaks structure} 
To determine the reason for this effect, simulations have been run using the AllPix~\cite{allpix} Geant4-based software.
The model used for the simulation reproduces the FE-I4 telescope: 6 planes, the two-mid planes rotated by $90^\circ$, sensors with FE-I4 geometry and $200\,\um$ thickness.
All planes were perfectly aligned to each other and the simulated beam of $180\,$GeV \textpi$^+$ had no angular spread.
Simulated data have been analysed in the same way as the real data using the Judith framework.

\newpage

\begin{figure}[!ht]
  \begin{center}
    \subfigure[]{
      \includegraphics[width=0.475\textwidth]{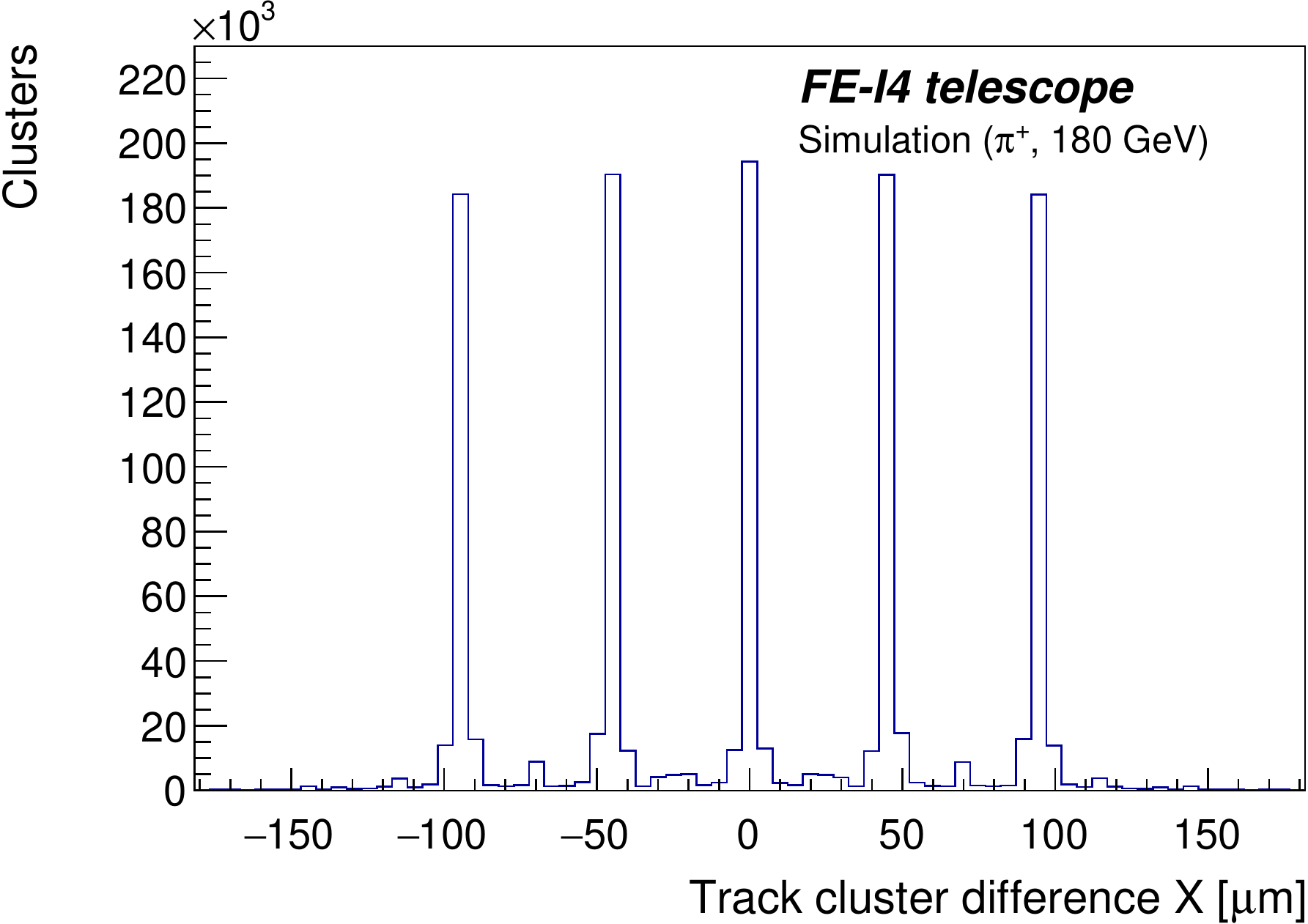}
      \label{fig:sim_residual}
    }
    \hfill
    \subfigure[]{
      \includegraphics[width=0.475\textwidth]{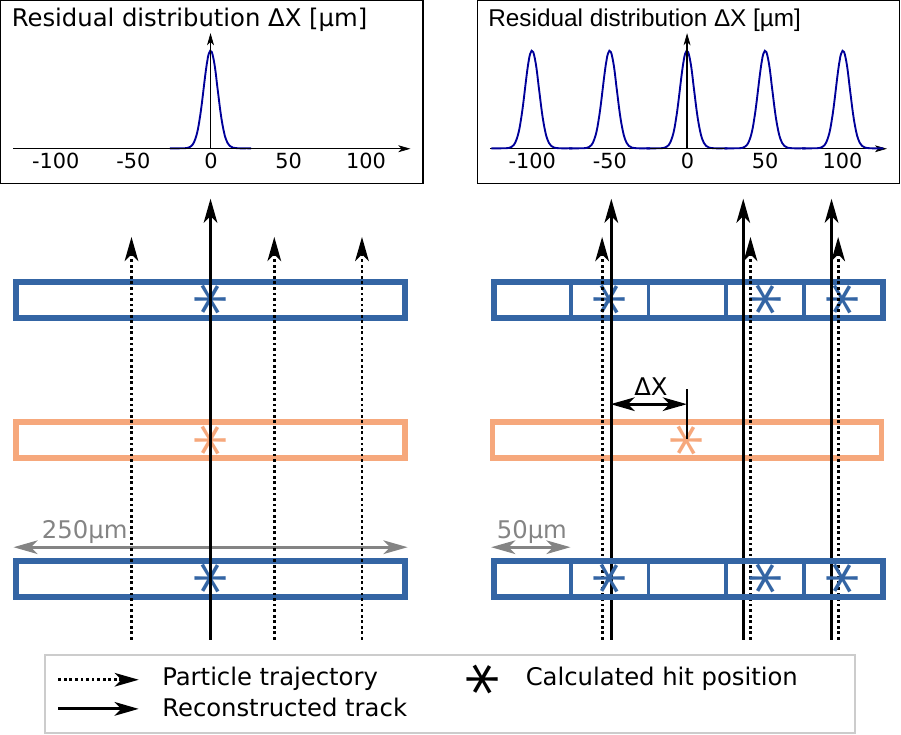}
      \label{fig:fivepeaks}
    }
    \vspace{-6pt}
    \caption[]{\subref{fig:sim_residual} Simulation of the residuals distribution for the third telescope plane in the X direction. All planes were perfectly aligned and no charge sharing between pixels was taken into account. \subref{fig:fivepeaks} Schematic description of the origin of the five peaks structure in the residuals distribution. The blue boxes represent pixels of tracking planes, while the orange box stands for a pixel of the DUT. The residual distribution is calculated as the distance between the reconstructed track position and the calculated hit position on the sensor.} 
  \end{center}
\end{figure}

As shown in Figure \ref{fig:sim_residual}, the residuals distribution along X, (the direction in which the pixel pitch is $250\,\um$) presents five peaks equally spaced by $50\,\um$.
Due to the simplified model the distribution is not disturbed by inefficiencies and misalignment of the detectors or charge sharing which would lead to broad and merging peaks like in Figure \ref{fig:residualX}.
The observed feature comes purely from the $250 \times 50\,\um^2$ rectangular shape of the pixels.
\newpage
For simplicity only single hit clusters and the simple model described above are considered.
Excluding the rotated planes, the reconstructed tracks would exactly follow the centres of the hit pixels as shown on the left hand side of Fig. \ref{fig:fivepeaks}, leading to a sharp residual distribution, centred at zero.
The rotated planes increase the granularity for tracking by $p_X/p_Y=5$ in the X direction, which enables to distinguish between tracks with a resolution of $50\,\um$ as depicted on the right hand side of Fig. \ref{fig:fivepeaks}.
This leads to five possible reconstructed track positions on the DUT and therefore to the observed effect.

In conclusion, this effect only influences the shape of the residuals tops, that are generally assumed to be flat.

  \section{Conclusions}\label{sec:sum}
      A telescope based on the IBL FE-I4 modules was constructed in 2014 and since then successfully  employed in various testbeam measurements at the CERN SPS H8 beam line.
It records tracks with a rate of around $18\,$kHz in stand alone mode and $6\,$kHz with up to two FE-I4 DUTs.
Upper limits on the resolution were calculated to be $11.7\,\um$ and $8.3\,\um$ in the XY plane at the DUT position.
The telescope allowed the testing of various detectors, ranging from small and full size FE-I4 based sensors~\cite{usr-higheta}\cite{h18tb} to strip modules and pad detectors~\cite{usr-lorenzo} read out by fully independent DAQ systems that were synchronised to the telescope by means of two trigger schemes.
In the case of very small devices the Region Of Interest trigger speeded up data taking by a factor of 35 compared to triggering on the full telescope acceptance.
Services providing power, cooling and monitoring are incorporated in the telescope setup and centrally managed by the user, allowing for a completely remote controlled data taking.
As a semi-permanent installation of a highly integrated system the telescope is available to support detector development at the CERN SPS beam lines.

  \acknowledgments
    We wish to thank the ATLAS IBL collaboration for providing the telescope's detector modules and gratefully acknowledge the support by the CERN PS and SPS instrumentation team.
We thank Allan Clark for a careful reading of the manuscript.
The research presented in this paper was supported by the SNSF grants 200020\_156083 and 20FL20\_160474.

\end{document}